\documentstyle[pre,preprint,aps,epsfig,amssymb]{revtex}
\tighten
\begin{document}

\title{Patterns arising from the interaction between scalar and vectorial 
instabilities in two-photon resonant Kerr cavities}

\author{Miguel Hoyuelos, Daniel Walgraef\cite{Daniel}, Pere Colet and 
Maxi San Miguel}

\address{Instituto Mediterr\'aneo de Estudios Avanzados\cite{www}, IMEDEA
(CSIC-UIB), \\
Campus Universitat Illes Balears, E-07071 Palma de Mallorca,
Spain.}

\date{January 27, 1999}

\maketitle

\begin{abstract} 
We study pattern formation associated with the polarization degree of freedom
of the electric field amplitude in a mean field model describing a nonlinear
Kerr medium close to a two-photon resonance, placed inside a ring cavity with 
flat mirrors and driven by a coherent $\hat x$-polarized plane-wave field. 
In the self-focusing case, for negative detunings the pattern arises naturally 
from a codimension two bifurcation. For a critical value of the field
intensity there are two wave numbers that become unstable simultaneously,
corresponding to two Turing-like instabilities. Considered alone, one of the
instabilities would originate a linearly polarized hexagonal pattern whereas
the other instability is of pure vectorial origin and would give rise to an
elliptically polarized stripe pattern. We show that the competition between the
two wavenumbers can originate different structures, being the detuning a
natural selection parameter.
\end{abstract}


\section{Introduction}

Spatiotemporal patterns in the transverse direction of an optical field have
now been widely studied theoretically and experimentally \cite{Chaossolfract}.
Studies of optical pattern formation share a number of aspects and techniques
with general investigations of pattern formation in other physical systems
\cite{CrossHoh}, but they also have specific features such as the role of light
diffraction and the vectorial degree of freedom associated with the
polarization of the light electric field amplitude. A prototype simple model
which has been very useful for the understanding of pattern formation in
nonlinear optical cavities is a mean field model describing a Kerr medium in a
cavity with flat mirrors and driven by a coherent plane-wave
field\cite{lugi,firt}. This model was extended to take into account the
polarization degrees of freedom in \cite{geddes1,corrkerr,hoyuelos}. Some of
the basic polarization mechanisms of pattern formation in alkali vapors or
other non-linear materials can be understood in terms of this simple model
despite the fact that the model is too simple to give a complete description of
alkali vapors. Furthermore, the relative simplicity of the model in
\cite{geddes1} makes it worthwhile to study it in depth as a general prototype
model for the basic understanding of vectorial patterns. A first study was
undertaken in \cite{geddes1} for the case of linearly polarized driving field
and the positive cavity detuning. A more detailed study in which the case of
elliptically polarized driving field is also considered has been presented
recently in reference \cite{hoyuelos}. 

Cavityless Kerr media can be used as an optical phase conjugation mirror. An
ideal phase conjugate mirror should generate an output field such that the
amplitude of the field, its propagation vector and its polarization unit 
vector are the complex conjugates of the corresponding magnitudes in the input
field. The first two properties can be achieved easily using, for example,
four-wave mixing processes \cite{Boyd92}. Usually the third property can be
obtained only when the pump waves are circularly polarized and counterrotating.
In order to find a system that verifies the three properties (vector phase 
conjugation) for an arbitrarily polarized input field it is required to make 
use of the special tensor properties of two-photon atomic transitions
in degenerate four-wave mixing processes \cite{Grynberg84,Ducloy84}. More
precisely, vector phase conjugation can be achieved if the two-levels coupled
by the two-photon transition have equal angular momenta $J$ with $J=0$ or
$J=1/2$. In this situation the $\chi_{1122}$ component of the susceptibility
tensor vanishes. 
Intuitively, as $\Delta J=0$, the atom do not change its angular momentum
either by absorbing two pump photons or by emitting a probe and a conjugate
photon so that the conjugate photon must be emitted with angular momentum equal
and opposite to that of the probe photon. A detailed calculation
\cite{Grynberg84} shows that indeed this is true for the cases indicated 
before. We should stress that the polarization properties of 
two-photon-resonant degenerate four-wave mixing processes are different from
those of most other degenerate four-wave mixing processes. In the two-photon 
case the underlying physical mechanism is scattering of the probe field from a 
spatially uniform temporally varying coherence induced by the two pump waves 
whereas in the other cases is scattering from a spatially varying 
refractive-index distribution induced by the interference between the pump 
and the prove beams.
Experimentally, vector phase conjugation was first observed using the $3
S_{1/2} \rightarrow 6 S_{1/2}$ two-photon transition in sodium vapor
\cite{Malcuit88}.

Here we will show that for a ring cavity filled with a nonlinear Kerr medium 
close to a two-photon-resonance (so that $\chi_{1122}=0$)
and illuminated with linearly polarized input and with negative detuning a new
interesting situation appears: A codimension two bifurcation in which two
stationary Turing-like instabilities occur simultaneously. The first
instability, if the other were not present, would originate a hexagonal pattern
which is polarized linearly in the same direction as the input field. The
second instability is of pure vectorial origin and if the first instability
were not present it would give rise to an elliptically polarized stripe
pattern. Here we study the interplay between the two instabilities. The
codimension two bifurcation appears here in a natural way associated to the
two-photon-resonant four-wave mixing nonlinearities rather that as the result
of the fine tuning of two system parameters as is usually the case. The
intensity of the pump field is the single control parameter to be tuned to
change the distance to the instabilities. Furthermore, the system still have
another easily accessible control parameter, the detuning, which allows the
system to form different patterns while remaining at the same distance to the
codimension two instability threshold. In particular we show how the detuning
can be used as a tuning parameter to select the pattern.

The outline of this paper is as follows: In Sec. \ref{model} we describe the
model we are considering, its spatially homogeneous solution and the stability
analysis of this solution. In Sec. \ref{weaknonlin} using a weakly nonlinear
analysis we derive the evolution equation for the patterns arising from the 
interacting instabilities. From these equations, the selected patterns for
different values of the detuning are analyzed in Sec. \ref{patternselection}.
In Sec. \ref{numerical} we describe the results from numerical simulations of
the model and finally in Sec. \ref{conclusions} we give some concluding
remarks.

\section{Description of the model, reference steady states and stability
analysis}
\label{model}

The system we consider is a ring cavity filled with an isotropic Kerr medium.
The cavity is driven by an external $\hat x$ polarized input field. The
situation in which the polarization degree of freedom of the electromagnetic
field is frozen was first considered by L. A. Lugiato and R. Lefever
\cite{lugi,firt}. Geddes et al \cite{geddes1} generalized the model of
\cite{lugi} to allow for the vector nature of the field. Their description of
this system is given by a pair of coupled equations for the evolution of the
two circularly polarized components of the field envelope $E_+$ and $E_-$,
defined by 
$$ 
E_\pm = \frac{1}{\sqrt{2}}(E_x \pm iE_y). 
$$ 
For an isotropic medium, the equations are
\begin{equation}
\frac{\partial E_\pm}{\partial t} = -(1 + i\eta \theta) E_\pm + 
i a \nabla^2 E_\pm + E_0 + i\eta [ A |E_\pm|^2 + (A+B) |E_\mp|^2] E_\pm ,
\label{1}
\end{equation}
where $E_0$ represents the components of the input field (the right and left
circularly polarized components are equal since we consider $\hat x$ linearly
polarized input), $\eta = +1 \ (-1)$ indicates self-focusing (self-defocusing),
$\theta$ is the cavity detuning, $a$ represents the strength of diffraction and
$\nabla^2$ is the transverse Laplacian. The parameters $A$ and $B$ are related
to the nonlinear susceptibility tensor components in the following way: 
$A=6\chi_{1122}$ and $B=6\chi_{1221}$ \cite{Boyd92}. Also, for an isotropic 
medium we have $A + B/2 = 1$. As discussed in the introduction, here we are
considering two-photon transitions between levels with equal angular momenta
$J$ where $J=0$ or $J=1/2$, so that $\chi_{1122}=0$ ($A=0$ and $B=2$). Also as 
we consider the self-focusing situation, in what follows we take $\eta =
+1$. The intensity of the input field is $I_0 = 2 |E_0|^2$.

\begin{figure}
\psfig{figure=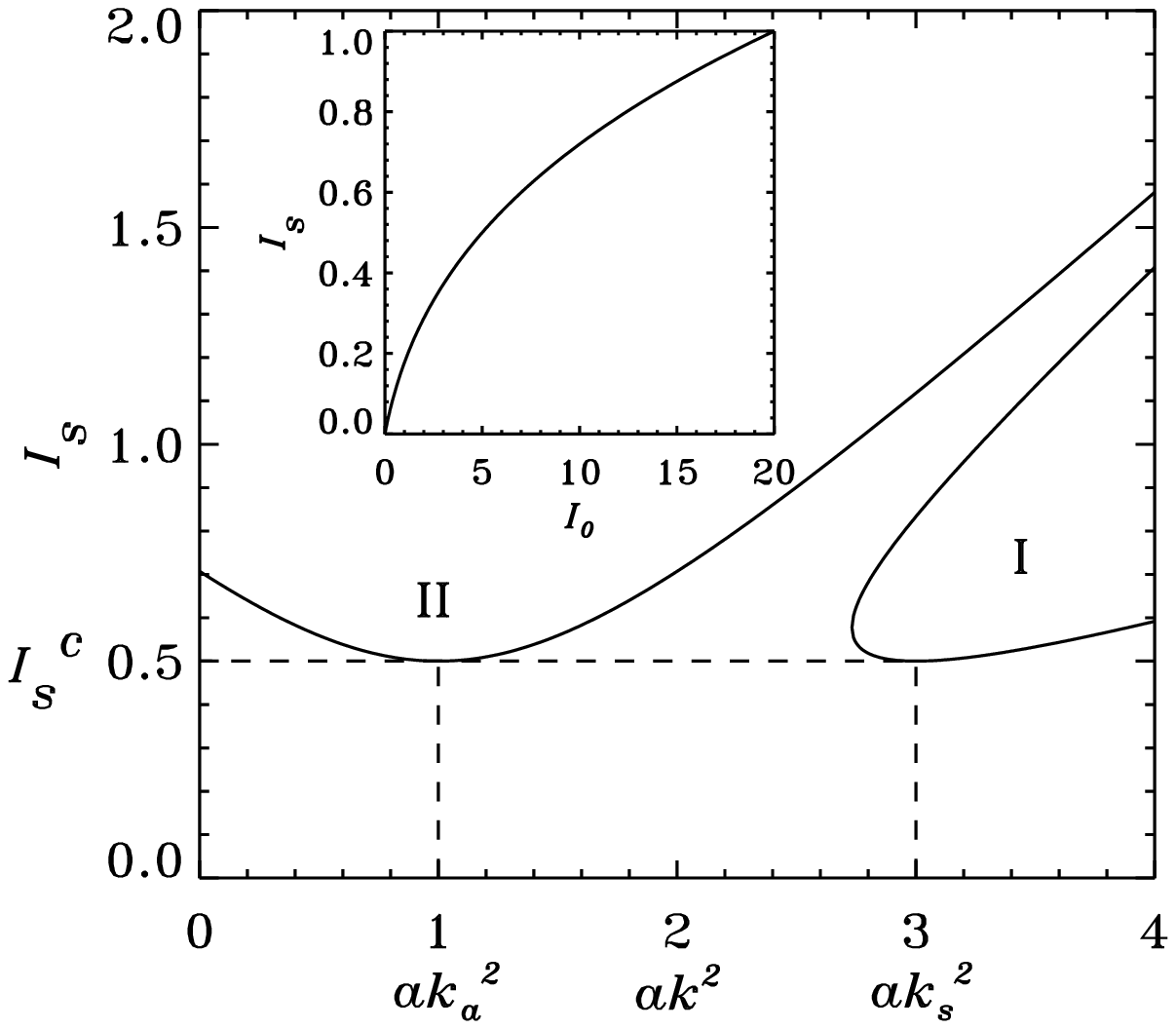,width=14cm}
\caption{Marginal stability curves for linearly polarized input field
corresponding to the symmetric solution. In the inset the symmetric steady
state homogeneous solution is shown, as a function of the input field
intensity, for linearly polarized light. Value of the detuning $\theta=-1$. The
parameters used in Eq. (\ref{1}) in order to get the codimension two situation
are: $\eta = 1$, $a=1$, $A=0$ and $B=2$. The quantities plotted in all the
figures are dimensionless.}
\label{fig1}
\end{figure}

The steady state homogeneous solutions of Eq. (\ref{1}) are reference states
from which transverse patterns emerge as they become unstable. There is a
symmetric ($E_{s+}=E_{s-}=E_s$) and two asymmetric
($E_{s+} \neq E_{s-}$) homogeneous solutions \cite{hoyuelos}. The symmetric
solution corresponds to linearly polarized output light, while the asymmetric
solutions to elliptically polarized output. Increasing the input field, 
the asymmetric solutions appear only for values of $I_0$ larger than the 
instability threshold for pattern formation\cite{hoyuelos}, so here we will 
only consider the symmetric solution\cite{footnote1},
\begin{equation}
I_0/2 = I_s(1 + (2I_s - \theta)^2),
\label{symmetricsol}
\end{equation}
which gives an implicit formula for $I_s=|E_s|^2$. As it is well known, Eq. 
(\ref{symmetricsol}) implies bistability for $\theta > \sqrt{3}$. However here
we will always consider negative detunings which are far away from the bistable
regime. An example of the symmetric solution for linearly polarized input is
given in the inset of Fig. \ref{fig1}. 

Basic features of the stability of the steady state homogeneous symmetric 
solution can be analyzed by considering the evolution equations for 
perturbations $\psi_{\pm}$ defined by
\begin{equation}
E_{\pm} = E_{s}(1+ \psi_{\pm}) \,.
\label{perturba}
\end{equation}
From Eqs. (\ref{1}) and (\ref{perturba}) we find
\begin{equation}\label{pertev}
\partial_t \psi_{\pm} = -[1+i (\theta - 2I_s) - ia\nabla^2]\psi_{\pm} 
+ i I_s 2(\psi_{\mp}+\psi_{\mp}^*+\vert\psi_{\mp}\vert^2)(1+\psi_{\pm}) \,.
\end{equation}

It is convenient to make a change of variables to the following basis 
\cite{geddes1}
\begin{equation}
\Sigma = \left(\sigma_1, \sigma_2, \sigma_3, \sigma_4 \right)^T =
\left(\Re(\psi_+ + \psi_-), \Im(\psi_+ + \psi_-), \Re(\psi_+ - \psi_-),
\Im(\psi_+ - \psi_-)\right)^T \,,
\end{equation}
where $T$ stands for transpose. In this basis, which emphasizes the role of
symmetric ($\psi_+=\psi_-$) and antisymmetric ($\psi_+=-\psi_-$) modes, Eq.
(\ref{pertev}) may be written as:
\begin{equation} \label{vectevol}
\partial_t \Sigma = L \Sigma + N_2(\Sigma\vert\Sigma) + 
N_3(\Sigma\vert\Sigma\vert\Sigma) ,
\end{equation}
where the linear matrix $L$ is a matrix with $2\times 2$ blocks in which the 
symmetric ($\sigma_1$, $\sigma_2$) and antisymmetric ($\sigma_3$, $\sigma_4$) 
modes are decoupled.
\begin{equation}
L = \left( \begin{matrix}
{L_1 & 0\cr 0 & L_2}\end{matrix} \right).
\end{equation}
As a consequence, the linear instabilities lead to the growth of either a
symmetric or an antisymmetric mode. In Fourier space we have:
\begin{equation}
L_1 = \left( \begin{matrix}
{-1 & (\theta - 2I_s + a k^2)\cr 
-(\theta - 6I_s +a k^2) & -1}\end{matrix} \right)
	\label{L1}
\end{equation}
and
\begin{equation}
L_2 = \left( \begin{matrix}
{-1 & (\theta - 2I_s + a k^2)\cr
 -(\theta +2I_s + a k^2) & -1}\end{matrix} \right),
	\label{L2}
\end{equation}
where $k \equiv |\vec k|$.

Instability occurs if at least one of the eigenvalues $\lambda$ of $L_1$ and
$L_2$ has a positive real part. In Fourier space, these eigenvalues are
solutions of the characteristic equations:
\begin{eqnarray}
(\lambda_1 +1)^2 &+&( \theta + a k^2 - 6I_s)(\theta + a k^2 - 2I_s)=0 \nonumber\\
(\lambda_2 +1)^2 &+&(\theta + a k^2 +2I_s)(\theta + a k^2 - 2I_s)=0 
\end{eqnarray}
For $\theta < 0$, we have 
\begin{eqnarray}
\lambda_{1\pm} &=& - 1 \pm \sqrt{4I_s^2 - a^2(k^2-k_s^2)^2}\nonumber \\
\lambda_{2\pm} &=& - 1 \pm \sqrt{4I_s^2 - a^2(k^2-k_a^2)^2} \,,
\label{lambdasymmetric}
\end{eqnarray}
where $ak_s^2 = 4I_s + \vert\theta\vert $ and $ak_a^2 =\vert\theta\vert$. For
both eigenvalues, the instability occurs at $I_s=1/2$.

The non linearities in (\ref{vectevol}) include quadratic 
$N_2(\Sigma\vert\Sigma)$ and cubic terms $N_3(\Sigma\vert\Sigma\vert\Sigma)$:
\begin{equation}
N_2(\Sigma\vert\Sigma) = I_s\left( \begin{matrix}
{2 \sigma_3 \sigma_4 - 2\sigma_1 \sigma_2\cr 
3\sigma_1 \sigma_1 + \sigma_2 \sigma_2 - \sigma_3 \sigma_3 + 
\sigma_4 \sigma_4 \cr 
2\sigma_2 \sigma_3 - 2\sigma_1 \sigma_4\cr 
-2\sigma_1 \sigma_3 - 2\sigma_2 \sigma_4}\end{matrix} \right) 
\end{equation}
and 
\begin{equation}
N_3(\Sigma\vert\Sigma\vert\Sigma) = {I_s\over 2}\left( \begin{matrix}
{2 \sigma_1 \sigma_3 \sigma_4 - \sigma_2(\sigma_1\sigma_1 + \sigma_2\sigma_2
- \sigma_3\sigma_3 + \sigma_4\sigma_4 )\cr 
\sigma_1 (\sigma_1 \sigma_1 + \sigma_2\sigma_2
+ \sigma_3\sigma_3 - \sigma_4\sigma_4 )-2 \sigma_2 \sigma_3 \sigma_4\cr 
2 \sigma_1 \sigma_2 \sigma_3 - \sigma_4(\sigma_1\sigma_1 - \sigma_2\sigma_2
+ \sigma_3\sigma_3 + \sigma_4\sigma_4 )\cr 
\sigma_3(-\sigma_1\sigma_1 + \sigma_2\sigma_2 + \sigma_3\sigma_3 + 
\sigma_4\sigma_4 ) - 2 \sigma_1 \sigma_2 \sigma_4}\end{matrix} \right) \,.
\end{equation}
The structure of these terms also gives some general information on the nature
of the instabilities. In particular, if the quadratic non linearity
$N_2(\Sigma\vert\Sigma)$ does not vanish, one expects the formation of
hexagonal patterns instead of stripes. As explained in Ref. \cite{hoyuelos}
when the symmetric mode becomes unstable an hexagonal pattern is expected
whereas when the antisymmetric mode becomes unstable there are no relevant
quadratic couplings so a stripe pattern is formed. 

In Fig. \ref{fig1} we plot marginal stability curves for $\theta=-1$ as a
function of $ak^2$. The shape of the marginal stability curves is, in fact, the
same for any value of the detuning $\theta$. This is because the eigenvalues
$\lambda_i$ given by Eq. (\ref{lambdasymmetric}) depend on $ak^2 - |\theta|$,
so a change in the value of $\theta$ is equivalent to a displacement of the
origin of $ak^2$ by the same amount. The origin moves to the right if the
detuning $\theta$ is increased.

The instability region I comes from the eigenvalue $\lambda_{1+}$ so the 
critical modes are symmetric and of zero frequency. A subcritical hexagonal 
pattern is expected via a transcritical bifurcation. If this were the only 
instability, it would correspond to the case discussed in \cite{firt}, in which
the polarization degree of freedom is not taken into account. This instability
leads to an $\hat x$-polarized pattern while the $\hat y$-polarized component
of the field continues to be zero. 

The instability region II comes from the eigenvalue $\lambda_{2+}$ so the
critical modes are antisymmetric and of zero frequency. A stripe pattern is
expected \cite{geddes1}. Given the antisymmetric nature of the unstable mode,
the $\hat x$-polarized component of the field is stable and remains almost
homogeneous, while the stripe pattern appears in the $\hat y$-polarized
component, which has zero value below the instability. Overall, the electric
field displays an elliptically polarized spatial structure. We remark that such
an instability is of pure vectorial nature with no analogue when the
polarization degree of freedom is frozen. 

In the case considered here, $\theta \le 0$, starting from the linearly
polarized homogeneous solution, as the input field is increased, the system
crosses the two instability thresholds simultaneously. This is a codimension
two bifurcation involving two sets of stationary modes. The critical modes
associated to region I are symmetric and have a critical wave number $k_s$, 
while the critical modes associated to region II are antisymmetric and have a 
critical wave number $k_a$. The ratio $k_a/k_s$ can be changed easily varying 
the value of $\theta$.

\section{Weakly Nonlinear Analysis for Interacting Turing Instabilities}
\label{weaknonlin}

The eigenmodes of the linear evolution matrix $L$ are, in Fourier
space $(\hat S_+(\vec k),0,0,0)^T$, $(0,\hat S_-(\vec k),0,0)^T$, 
$(0,0,\hat A_+(\vec k),0)^T$ and $(0,0,0,\hat A_-(\vec k))^T$, with
\begin{equation}
\hat S_{\pm}(\vec k) = \hat\sigma_1(\vec k) \pm \beta_s (k)\hat\sigma_2(\vec
k), \quad
\hat A_{\pm}(\vec k) = \hat\sigma_3(\vec k) \mp \beta_a (k)\hat\sigma_4(\vec
k),
\end{equation}
where $\hat U(\vec k)={\cal F} [U](\vec k)$ denotes the Fourier transform of
$U(\vec r)$. Furthermore
\begin{equation}
\beta_j (k)= {2I_s + a(k^2-k_j^2)\over \sqrt{4I_s^2 - a^2(k^2-k_j^2)^2}} \, ,
\end{equation}
where index $j$ stands for $s$ or $a$ and $\beta_j (k_j) = 1$. The critical 
modes correspond to the eigenvalues $\lambda_{1+}(k_s)$ and $\lambda_{2+}(k_a)$.
Note also that
\begin{equation}
\hat\sigma_1 = {\hat S_+ + \hat S_- \over 2 } \, , \quad
\hat\sigma_2 = {\hat S_+ - \hat S_- \over 2\beta_s} \, , \quad
\hat\sigma_3 = {\hat A_+ + \hat A_- \over 2 } \, , \quad
\hat\sigma_4 = -{\hat A_+ - \hat A_- \over 2\beta_a } \, .
\end{equation}

After diagonalization of the linear evolution matrix, the dynamics (\ref
{vectevol}) may be rewritten, in Fourier space, as
\begin{equation}
\partial_t \hat S(\vec k) = \hat\Lambda(\vec k) \hat S(\vec k)
 + \hat N_2(\hat S\vert\hat S)\vert_{\vec k} + 
 \hat N_3 (\hat S\vert\hat S\vert\hat S)\vert_{\vec k} \, ,
\end{equation}
where $\hat S = (\hat S_+, \hat S_-, \hat A_+, \hat A_-)^T$ and
$\hat\Lambda$ is a diagonal matrix with diagonal elements 
$\left\{\lambda_{1+},\lambda_{1-},\lambda_{2+},\lambda_{2-}\right\}$.
$\hat N_2\vert_{\vec k}$ and $\hat N_3\vert_{\vec k}$ are the Fourier
transforms of the nonlinear terms of equation (\ref{vectevol}), where the
$\sigma_i$ have been replaced by the corresponding linear combinations of
$S_+$, $S_-$, $A_+$ and $A_-$. Slightly above threshold, this dynamics may be
reduced to the dynamics of the critical modes amplitudes only, through the
adiabatic elimination of the stable non-critical modes. This procedure is now
standard \cite{daniel}, and we will only sketch here the main steps of its
application to model (\ref{vectevol}), and derive evolution equations for the
critical modes, up to cubic nonlinearities.

Let us write the amplitudes of the critical modes as $S(\vec k)\equiv
\hat S_+(\vec k)\delta(|\vec k|-k_s)$ and $A(\vec k ) \equiv \hat A_+(\vec k)
\delta(|\vec k|-k_a)$. Their dynamics writes
 \begin{eqnarray}
 \dot S(\vec k) &=& (2I_s - 1) S(\vec k) + 
 (\hat N_2\vert_{\vec k,k=k_s})_S + 
 (\hat N_3 \vert_{\vec k,k=k_s})_S \, ,\nonumber\\
 \dot A(\vec k) &=& (2I_s - 1) A(\vec k) +
 (\hat N_2\vert_{\vec k,k=k_a})_A + 
 (\hat N_3 \vert_{\vec k,k=k_a})_A \, .
 \end{eqnarray}
The terms in the quadratic nonlinearities $\hat N_2$ are convolutions of 
products of critical and non-critical modes. The terms involving non-critical
modes only ($\hat S_-^2$, $\hat A_-^2$, $\hat S_- \hat A_-$) may be neglected, 
since they contribute, through the adiabatic elimination process, to
nonlinearities of quartic or higher order. Furthermore, the structure of the
cubic nonlinearities of Eq. (\ref{vectevol}) is such that the pure critical
mode contribution to $\hat N_3$ vanishes. Hence $\hat N_3$ will be neglected.
We are thus dealing here with a non-generic case, where cubic nonlinearities
are generated solely by the adiabatic elimination of stable modes from
quadratic terms. The nonlinearities $(\hat N_2\vert_{\vec k,k=k_s})_S$ and
$(\hat N_2\vert_{\vec k,k=k_s})_A$ can be written at the lowest order in a
non-critical modes as
\begin{eqnarray}
(\hat N_2\vert_{\vec k,k=k_s})_S&\simeq& {I_s\over 2}
\int{ d\vec k' \left[ S(\vec k - \vec k')S(\vec k') - A(\vec k - \vec k')
A(\vec k') \right. } \nonumber\\ 
&& \quad \quad \left. + 2S(\vec k - \vec k') \hat S_-(\vec k') 
\delta(|\vec k'|-k_s) - 2A(\vec k-\vec k') \hat A_-(\vec k') 
\delta(|\vec k'|-k_a)\right]\nonumber\\ 
&&+ I_s \left[ S(\vec k) (\hat S_+(0) + \hat S_-(0)) + 
S(-\vec k)(\hat S_+(2\vec k) + \hat S_-(2\vec k))\right] \, ,\nonumber\\
(\hat N_2\vert_{\vec k,k=k_a})_A &\simeq& I_s \int{d\vec k' 
\left[ S(\vec k - \vec k') A(\vec k') \right.} \nonumber\\
&& \quad \quad \left. + S(\vec k-\vec k')\hat A_- (\vec k') 
\delta(|\vec k'|-k_a) + A(\vec k-\vec k')\hat S_-(\vec k') 
\delta(|\vec k'|-k_s) \right] \nonumber\\
&&+ I_s A(\vec k)\left[\hat S_+(0) +\hat S_-(0)\right] \, .
\label{n2sn2a} 
\end{eqnarray}
The non-critical modes $\hat S_{\pm}(0)$, $\hat S_{\pm}(2\vec k_s)$, $\hat
S_-(\vec k_s)$ and $\hat A_-(\vec k_a)$ present in the Eqs. (\ref{n2sn2a}) may
be expressed as an expansion in powers of critical ones using the adiabatic
elimination procedure. One has at the leading order
\begin{eqnarray}
\hat S_{\pm}(0) &=& {I_s\over 2\lambda_{1,\pm }(0)}\int{d\vec k \left[A(\vec
k)A(-\vec k)+ (1 \mp 2 \beta_s (0))S(\vec k)S(-\vec k)\right]}\, , \nonumber\\
\hat S_{\pm}(2\vec k_s) &=& {I_s\over 2\lambda_{1,\pm}(2k_s)}(1 \mp 2 \beta_s
(2k_s)) S(\vec k_s)S(\vec k_s) \, ,\nonumber\\
\hat S_-(\vec k_s) &=& {I_s\over 2\lambda_{1,-}(k_s)}\int{d\vec k'\left[
3 S(\vec k_s-\vec k')S(\vec k')+A(\vec k_s-\vec k')A(\vec k')\right]} \, ,
\nonumber\\
\hat A_-(\vec k_a) &=& -{I_s\over \lambda_{2,-}(k_a)}\int{d\vec k'
S(\vec k_a-\vec k')A(\vec k')} \, .
\label{noncriticalmodes}
\end{eqnarray}
The substitution of Eqs. (\ref{noncriticalmodes}) in Eqs. (\ref{n2sn2a}) leads 
to the following asymptotic dynamics for the critical modes, valid close to 
the instability threshold
\begin{eqnarray}\label{noresampl}
\dot S(\vec k) &=& (2I_s - 1) S(\vec k) + \int{d\vec k' 
\left[v_0 S(\vec k - \vec k' )S(\vec k') + v_1 
A(\vec k - \vec k' )A(\vec k')\right]} 
\nonumber\\ 
&&- \int{d\vec k'\int{d\vec k'' u(\vec k,\vec k',\vec k'')
S(\vec k-\vec k')S(\vec k'-\vec k'') S(\vec k'')}} \nonumber\\ 
&&- \int{d\vec k'\int{d\vec k'' w(\vec k,\vec k',\vec k'') 
S(\vec k-\vec k')A(\vec k'-\vec k'') A(\vec k'')}}
\nonumber\\
\dot A(\vec k)&=& (2I_s - 1) A (\vec k)+ 
v_2 \int{d\vec k' S(\vec k - \vec k')A(\vec k')} \nonumber\\ 
&&- \int{d\vec k'\int{d\vec k'' \bar w(\vec k,\vec k',\vec k'')
A(\vec k-\vec k')A(\vec k'-\vec k'') A(\vec k'')}} \nonumber\\ 
&&- \int{d\vec k'\int{d\vec k'' \bar u(\vec k,\vec k',\vec k'') 
A(\vec k-\vec k')S(\vec k'-\vec k'') S(\vec k'')}}\, ,
\end{eqnarray}
where $v_0 = -v_1 = {v_2\over 2} = {I_s\over 2}$, and
\begin{eqnarray}
u(\vec k,\vec k',\vec k'') &=& u_1 \delta (|\vec k'| - k_s)
+ u_2 \delta(\vec k')
+ u_3 \delta(\vec k')\delta(\vec k''-\vec k) 
\nonumber \\
w(\vec k,\vec k',\vec k'') &=& w_1 \delta (|\vec k'| - k_s)
+ w_2 \delta(\vec k')
+ w_3 \delta (\vert \vec k-\vec k'-\vec k''\vert - k_a) 
\nonumber \\
\bar u(\vec k,\vec k',\vec k'') &=& u_1 \delta (|\vec k'| - k_s)
+ u_2 \delta(\vec k')
- w_3 \delta (\vert \vec k-\vec k'\vert - k_a)
\nonumber \\
\bar w(\vec k,\vec k',\vec k'') &=& w_1 \delta (|\vec k'| - k_s) 
+w_2 \delta(\vec k') \, ,
\end{eqnarray}
with $u_1={3I_s^2 \over 2+4I_s}$,
$u_2=I_s^2 {3+2|\theta| \over (2+|\theta|)^2}$,
$u_3=-{I_s^2\over9}{13+6|\theta| \over (2+|\theta|)^2}$,
$w_1=w_3={I_s^2 \over 2+4I_s}$, and
$w_2={I_s^2 \over (2+|\theta|)^2}$.
It has to be noted that the $u$ terms come from quadratic
resonances between critical and non critical symmetric modes, while the $w$
terms come from quadratic resonances between symmetric and antisymmetric modes.
As in the case of isolated Turinglike instabilities, there is no quadratic 
resonance between critical antisymmetric modes only. As a result, pattern
formation is expected to strongly depend on the existence of quadratic
resonances between symmetric and antisymmetric modes. Hence, for the sake of
simplicity, we will consider separately the case with quadratic resonance
between symmetric modes only, and the case with quadratic couplings between
antisymmetric and symmetric modes, which is more intricate. In the latter case,
quadratic couplings are such that $\vec k_1 = \vec k_2 +\vec k_3 $, with 
$\vert \vec k_1\vert = k_s$ and $\vert \vec k_{2,3}\vert = k_a$. Since $\cos
\phi = {k_s \over 2k_a}$, where $\phi$ is the angle between symmetric and
antisymmetric vectors, quadratic resonance arise only for $\vert \theta\vert >
{4I_s \over 3}$.
 
\section{Pattern Selection and Stability}
\label{patternselection}

In this section we study the various patterns that may appear as asymptotic
solutions of eq. (\ref{noresampl}), and their stability. Each of these patterns
is built on an arbitrary number of critical modes pairs, and it is the nature
of their nonlinear couplings that determines their stability. In the
following, we label the modes contributing to the formation of a pattern
$S(\vec k_i)=S_i$, $A(\vec k_i)=A_i$. We also use the notation $\bar S_i$ for
the complex conjugate of $S_i$.
 
\subsection{$k_a < {k_s \over 2}$ or $\vert \theta\vert < {4I_s \over 3}$}
\label{no-sym-antisym-coupling}

In this case, there is no quadratic coupling between symmetric and
antisymmetric modes, and there is no contribution coming form the terms with 
the coefficients $v_1$, $v_2$, $w_1$ and $w_3$. Let us then consider 
separately amplitude equations for each type of modes.
 
\subsubsection{Antisymmetric modes.}
For patterns built on antisymmetric modes only, the amplitude equations for an
arbitrary number $m$ of pairs of modes are:
\begin{equation}
\dot A_i = (2I_s - 1) A_i - 2 w_2 A_i\sum_{j=1}^m \vert A_j\vert^2 \, .
\end{equation}
Hence, a pattern built on $m$ pairs of wavevectors is marginally stable versus
a $m+1$ pair of wavevectors. So, at this level of analysis, any pattern with
an arbitrary number of wavevectors is possible, including patterns of the form
$A J_0(k_a r)$, where $J_0$ is zeroth order Bessel function and
$A=\sqrt{(2I_s-1)\over 2w_2}$. 
 
The linear growth rate of the evolution of symmetric modes in the presence of
such patterns is zero, so antisymmetric patterns are marginally stable versus
symmetric mode patterns. It would be necessary to go to higher orders in the
amplitude equations to complete the pattern selection analysis in this case.

\subsubsection{Symmetric modes.}
On the other hand, for patterns built on symmetric modes only, the amplitude
equations for a triplet of such modes are: 
\begin{equation}
\dot S_i = (2I_s - 1) S_i + 2 v_0 \bar S_{i+1}\bar S_{i-1}
- (2 u_2 + u_3)\vert S_i\vert ^2S_i
- 2 (u_1+u_2) S_i\sum_{j\ne i} \vert S_j\vert^2\, .
\end{equation}
At the instability threshold ($I_s={1\over 2}$), hexagonal pattern appear via a
subcritical bifurcation. Increasing the value of $I_s$, stripes may also become
stable for $2I_s-1 \ge {8(3+2\vert\theta\vert)\over 9(2+\vert\theta\vert)^2}$.
There is a small region of bistability of stripes and hexagons and for larger
values of $I_s$ only the stripes remain stable.

Hexagonal patterns are stable versus antisymmetric modes for $I_s \le {1\over
2}\left[1 + {(3+2\vert\theta\vert )\over 4(2+\vert\theta\vert )^2}\right]$. 
Hence, sufficiently close to threshold, one may expect hexagonal patterns.
On increasing $I_s$, such patterns should become unstable.

\subsubsection{Mixed modes.}
The amplitude equations for mixed structure formed by triplets of symmetric
modes and an arbitrary number of antisymmetric modes are:
\begin{eqnarray}
\dot S_i &=& (2I_s - 1) S_i + 2 v_0 S_{i+1}S_{i-1}\nonumber\\ 
&&- S_i \left[ (2u_2+u_3) \vert S_i\vert^2 + 2(u_1+u_2)
\sum_{j=1}^3 \vert S_j\vert^2 + 2 w_2 
\sum_{l=1}^m \vert A_l\vert^2 \right] \nonumber\\
\dot A_k&=& (2I_s - 1) A_k - A_k
\left[ 2 u_2\sum_{j=1}^3 \vert S_j\vert^2 + 2 w_2
\sum_{l=1}^m \vert A_l\vert^2 \right] \,.
\end{eqnarray}
However, these equations do not admit non trivial steady states. As a result,
in these conditions, hexagonal or striped patterns of symmetric modes, and
patterns built on an arbitrary number of antisymmetric modes may be
simultaneously stable.
 
\subsection{${k_s \over 2}<k_a < k_s$ or $\vert\theta\vert > {4I_s \over 3}$}
 
In this case, quadratic resonances may occur between symmetric and
antisymmetric modes, and one may now expect contributions coming from the 
$v_1$, $v_2$, $w_1$ and $w_3$ in the amplitude equations, which have to be
modified accordingly. Let us then consider the different types of patterns that
may arise in this case, and which are built on modes belonging to the following
set of critical modes (up to an arbitrary phase angle) (cf. figure
\ref{acopl}). 

\subsubsection{Symmetric modes.}
Striped and hexagonal symmetric modes patterns are now always unstable versus
antisymmetric modes. This is due to the positive nonlinear renormalization
induced by these modes in the evolution equation for their resonantly coupled
antisymmetric modes (this renormalization is associated to the terms with
$-w_3$ coefficients in $\bar u$).

\subsubsection{Antisymmetric modes.}
As in subsection \ref{no-sym-antisym-coupling}, a pattern built on $m$ 
arbitrary pairs of antisymmetric modes is marginally stable versus a $m+1$ 
pair of wavevectors. Furthermore, since the contributions of the quadratically
resonant symmetric and antisymmetric modes, in their respective amplitude
equations have opposite signs ($v_1<0$ and $v_2>0$), pure antisymmetric mode
patterns are also unstable versus resonantly coupled symmetric and
antisymmetric ones. Recall also that a pattern built on $m$ arbitrary pairs of
antisymmetric modes is marginally stable versus non resonant symmetric modes.

\subsubsection{Mixed modes.}
As a result, when quadratic resonances between symmetric and antisymmetric
modes are possible, pure steady patterns built on symmetric or antisymmetric
modes only, are always unstable. We have thus to consider the possibility for
the system to develop mixed modes patterns.

\begin{figure}
\vspace*{12cm}
\includegraphics{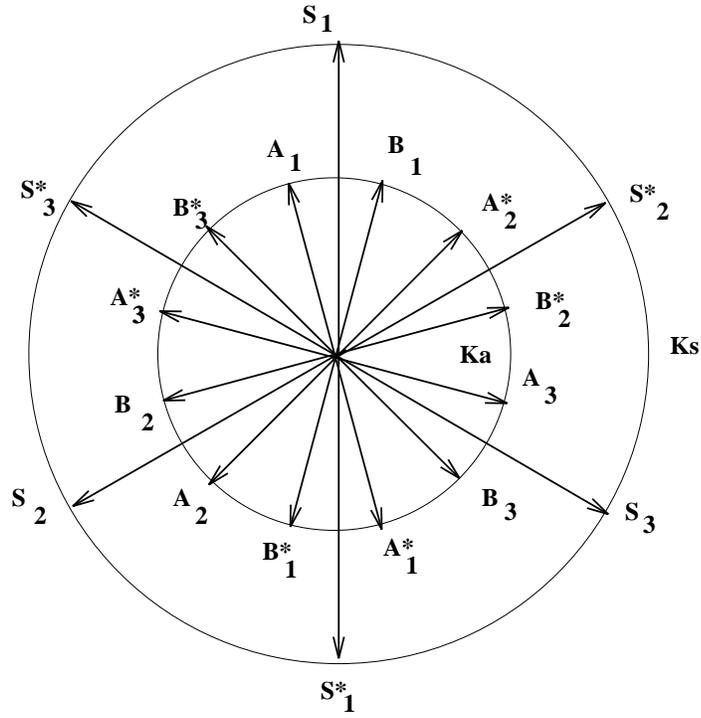} 
\caption{Quadratically coupled symmetric and antisymmetric modes with
wavenumbers $k_s$ and $k_a$}
\label{acopl}
\end{figure}

Let us consider the simplest case of mixed mode patterns built on one symmetric
mode and two quadratically resonant antisymmetric modes (e.g. $S_1$, $A_1$ and
$B_1$ in figure \ref{acopl}). Their uniform amplitude equations are:
\begin{eqnarray}
\dot S_1 &=& (2I_s - 1) S_1 - I_s A_1 B_1
- S_1 \left[ (2u_2+u_3)\vert S_1\vert^2 + 2w_2(\vert A_1\vert^2 + 
\vert B_1\vert^2) \right] \nonumber\\
\dot A_1&=& (2I_s - 1) A_1 + I_s S_1 \bar B_1 
- A_1 \left[ 2(u_2-w_3)\vert S_1\vert^2 + (2w_2\vert A_1\vert^2
+ (2w_2+w_1)\vert B_1\vert^2) \right] \nonumber\\
\dot B_1&=& (2I_s - 1) B_1 + I_s S_1 \bar A_1 
- B_1 \left[ 2(u_2-w_3)\vert S_1\vert^2 + (2w_2\vert B_1\vert^2 + 
(2w_2+w_1)\vert A_1\vert^2) \right] \, .
\end{eqnarray}
Because of the symmetry between $A_1$ and $B_1$ we look for solutions with
the same amplitude for both antisymmetric modes. 
Defining amplitude and phase variables as $S_1 = R_s\exp i\phi$, $A_1 =
R_a \exp i\psi $, $B_1 = R_a \exp i \bar\psi$, and 
$\Psi = \phi -\psi - \bar\psi$, one has
\begin{eqnarray}
\dot R_s &=& (2I_s - 1) R_s - I_s R_a^2\cos \Psi - 
R_s \left[ (2u_2 +u_3)R_s^2 + 4w_2R_a^2 \right]\nonumber\\
\dot R_a &=& (2I_s - 1) R_a + I_s R_aR_s\cos \Psi - 
R_a \left[ 2(u_2-w_3) R_s^2 + (4w_2+2w_1)R_a^2 \right] \nonumber \\
\dot \Psi&=& {I_s\over R_s} [R_a^2-2R_s^2]\sin\Psi \, .
\end{eqnarray}
A phase stable steady state corresponds thus to $\Psi = 2n\pi$ if $R_a^2
<2R_s^2$, and to $\Psi = (2n+1)\pi$ if $R_a^2 >2R_s^2$. Combining the steady
state conditions for $R_s$ and $R_a$, it can be seen that $\Psi = 2n\pi$
requires that $R_a^2 >R_s^2$ and that $\Psi = (2n + 1)\pi$ requires that $R_a^2
<R_s^2$. As a result, a stable steady state may only be obtained for $\Psi =
2n\pi$, with $R_s^2 < R_a^2 < 2R_s^2$. Furthermore, this condition is
satisfied if the kinetic coefficients are such that $1\le \vert \theta\vert
<2(\sqrt{3}-1)\simeq 1.46$, which is thus a necessary condition to be satisfied
to obtain such mixed modes solutions. This condition corresponds to $0.57 k_s <
k_a < 0.65 k_s$ or $0.64<\cos \phi < 0.769 $, with $I_s\simeq 0.5$.
 
A similar analysis may be performed for a pattern formed by an hexagonal 
planform of symmetric modes and their quadratically resonant antisymmetric 
ones (cf. fig \ref{acopl}). However, in this case, the fact that the quadratic
couplings between symmetric and antisymmetric modes have opposite signs does
not allow the stabilization of critical patterns.

\subsection{Summary of the analytical results.}

The conclusion of the analysis presented is the previous two subsections is 
as follows: 
\begin{itemize}
\item For $ k_a < 0.5 k_s$ ($\vert \theta\vert <0.666$), close to threshold 
hexagonal symmetric patterns are stable. For slightly larger values of $I_s$
symmetric stripes and hexagons are bistable and finally only the stripes remain
stable. Patterns built on an arbitrary number of antisymmetric modes are 
neutrally stable. As this results come from an expansion up to cubic
nonlinearities we can not conclude about the stability of the antisymmetric
patterns. It would be necessary to go to higher order terms. Finally, there are
no mixed stationary patterns.

\item For $0.5 k_s < k_a < 0.57 k_s$ ($0.666<\vert \theta\vert <0.963$),
pure symmetric and antisymmetric mode patterns are unstable. Also, no steady 
mixed mode patterns are found either. For that reason one expects to find
time dependent structures which involve both symmetric as well as antisymmetric
modes.

\item For $0.57 k_s < k_a < 0.65 k_s$ ($0.963<\vert \theta\vert < 1.46$), one
may expect steady patterns formed by the superposition of one symmetric mode 
and its quadratically resonant antisymmetric ones. We should notice that the 
range in which this steady mixed patterns does exist may be in fact smaller 
than $0.963<\vert \theta\vert < 1.46$. This is because our calculation was
for a necessary condition for the stability of the global phase, but this
is not a sufficient condition for the stability of the pattern.
 
\item For $0.65 k_s \le k_a$ ($1.46<\vert \theta\vert $), symmetric modes
are unstable versus antisymmetric modes and no steady mixed mode solutions are 
found. Patterns composed of an arbitrary number of antisymmetric modes are 
neutrally stable, so it would be necessary to include higher order
nonlinearities to conclude about their stability.
\end{itemize}
 
The present analysis provides the basic elements for the study of pattern
formation when both symmetric and antisymmetric modes become simultaneously
unstable. As it will be discussed below, it is partially confirmed by numerical
analysis of the complete dynamical model. It could be improved, on the one
hand, in determining the full stability range of mixed mode patterns, and, on
the other hand, in resolving the issue of finding asymptotic states in cases
where no steady critical patterns are found. In such cases, as suggested by
numerical analysis, one should consider the possibility of asymptotic
time-dependent or non critical patterns. The latter case would require to
incorporate in the dynamics harmonics or non critical modes.
We should notice that a peculiarity 
of the situation considered here is that there are no cubic nonlinear terms 
involving only critical modes. The cubic nonlinearities we have considered are 
generated solely by the adiabatic elimination of stable modes from the 
quadratic terms. However for an input field which is slightly above threshold 
there will be a range of unstable modes around the critical one. The cubic
nonlinearity $\hat N_3$ for these modes will be small but non zero. As the cubic 
nonlinearities generated by the adiabatic elimination of stable modes from 
the quadratic terms are also small, it would be necessary to include in the
analysis both kinds of cubic nonlinearities. This is beyond the scope of the 
present paper.

\section{Numerical results}
\label{numerical}

We have performed several numerical integrations of Eq. \ref{1} using a
numerical scheme described in detail in Ref. \cite{montagne}. The method is
pseudospectral and second-order accurate in time, and is similar to the
so-called two-step method. Lattices of size $128\times 128$ were used.

\begin{figure}
\psfig{figure=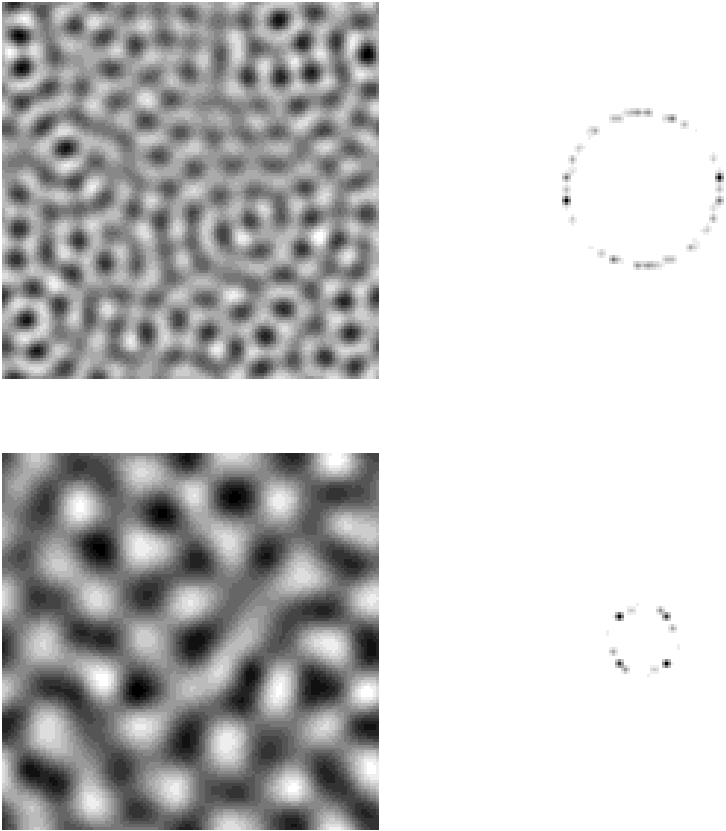}
\vspace*{1cm}
\caption{Field configuration for $\theta = -0.5$ and $I_s=0.51$ ($I_0=3.31$) 
after integrating Eq. \ref{1} for a time $t=35000$. From left to right and from
top to bottom: $\Re(E_x(\vec r))$ (near field $\hat x$-polarized component
plotted with the grayscale: black=0.51, white=0.60), $|E_x(\vec k)|^2$ (far
field $\hat x$ component), $\Re(E_y(\vec r))$ (near field $\hat y$-polarized
component plotted with the grayscale: black=-0.035, white=0.042) and $|E_y(\vec
k)|^2$ (far field $\hat y$ component).} 
\label{t05}
\end{figure}

For $\vert \theta \vert < 0.666$ there is not a mixed mode pattern since
quadratic resonance is not possible due to the fact that $k_a < k_s/2$. An
example of this situation is shown in fig. \ref{t05} for $\theta = -0.5$.
According to the weakly nonlinear analysis stationary symmetric hexagons are 
stable close to threshold, however we find numerically a disordered
structure for both polarization components of the field $E_x$ and $E_y$. As
$E_x=(E_+ + E_-)/\sqrt{2}$ and $E_y=(E_+ - E_-)/\sqrt{2}i$, the structure
observed in $E_x$ is formed by symmetric modes with wavevectors in the ring
$\vert \vec k \vert=k_s$ whereas the structure in $E_y$ is composed of
antisymmetric modes with wavevectors $\vert \vec k \vert=k_a$. The overall
structure is built on an arbitrary number of symmetric and antisymmetric modes.
The weakly nonlinear analysis predicted that there were no stationary mixed
patterns and, in fact, the pattern shown in fig. \ref{t05} from numerical
integration is not static, it evolves dynamically in a slow time scale.

\begin{figure}
\psfig{figure=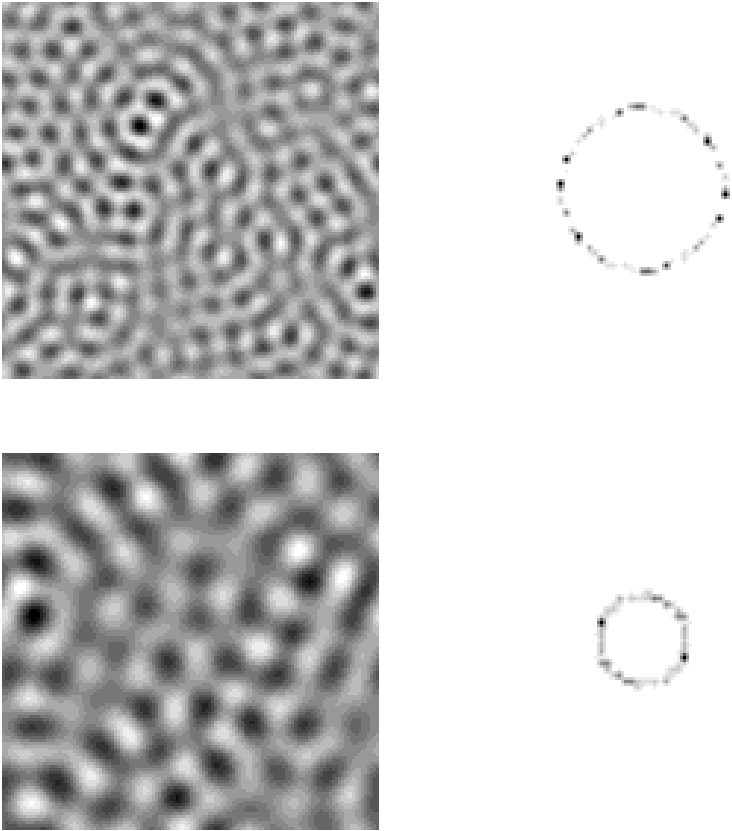}
\vspace*{1cm}
\caption{Field configuration for $\theta = -0.83$, $I_s=0.51$ ($I_0=4.41$) and
an integration time of $t = 65000$. From left to right and from top to bottom:
$\Re(E_x(\vec r))$ (grayscale: black=0.44, white=0.51), $|E_x(\vec k)|^2$,
$\Re(E_y(\vec r))$ (grayscale: black=-0.036, white=0.033) and $|E_y(\vec k)|^2$.}
\label{t83}
\end{figure}

For $0.666 < \vert \theta \vert < 0.963$, although resonance is possible, no
mixed mode pattern is numerically found, as predicted analytically. As in the
previous case, a disordered structure is obtained and no steady pattern is
reached even for long time integrations. Rings of radius $k_s$ and $k_a$ in the
far fields of the linearly polarized components $\hat x$ and $\hat y$
respectively, show that there is not a selection process of critical wave
vectors, all directions become unstable. In fig. \ref{t83} we show the field
configurations for $\theta = -0.83$. 

\begin{figure}
\psfig{figure=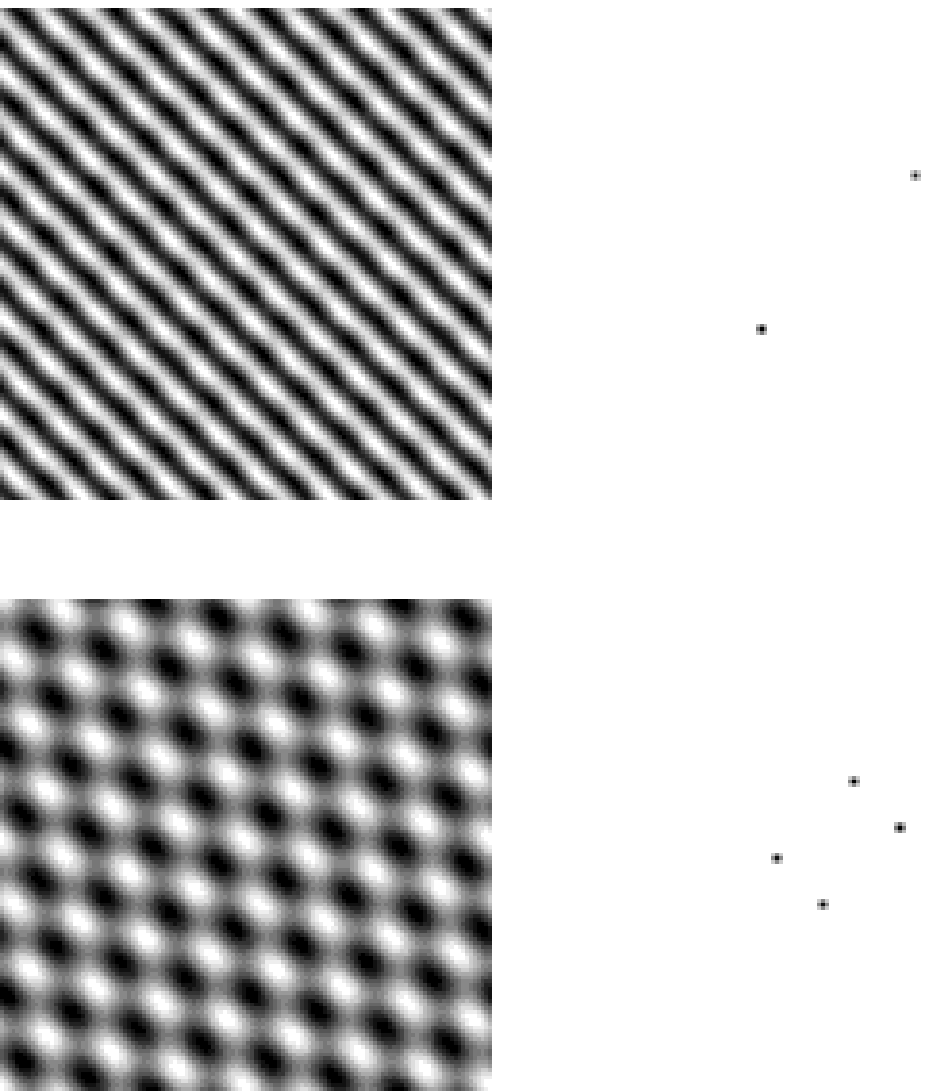}
\vspace*{1cm}
\caption{Field configuration for $\theta = -1$, $I_s=0.51$ ($I_0=5.06$) and an
integration time of $t = 14000$. From left to right and from top to bottom:
$\Re(E_x(\vec r))$ (grayscale: black=0.43, white=0.47), $|E_x(\vec k)|^2$,
$\Re(E_y(\vec r))$ (grayscale: black=-0.047, white=0.047) and $|E_y(\vec k)|^2$.}
\label{t1}
\end{figure}

The analysis of the previous section showed that mixed mode patters may be
expected for $0.963 < \vert \theta \vert < 1.46$. In fact, for $\theta = -1$ a pattern
formed by the superposition of a symmetric mode and the quadratically resonant
antisymmetric ones is obtained, see fig. \ref{t1}. A roll pattern is seen in
the $\hat x$ linearly polarized component with wave number $k_s$, while a
rectangular pattern appears in the $\hat y$ component with wave number $k_a$.
As for the range of existence of mixed mode patterns, we have found 
numerically the existence this patterns also for $\theta = -1.05$ but not
for $\vert \theta \vert \geq 1.1$.

\begin{figure}
\psfig{figure=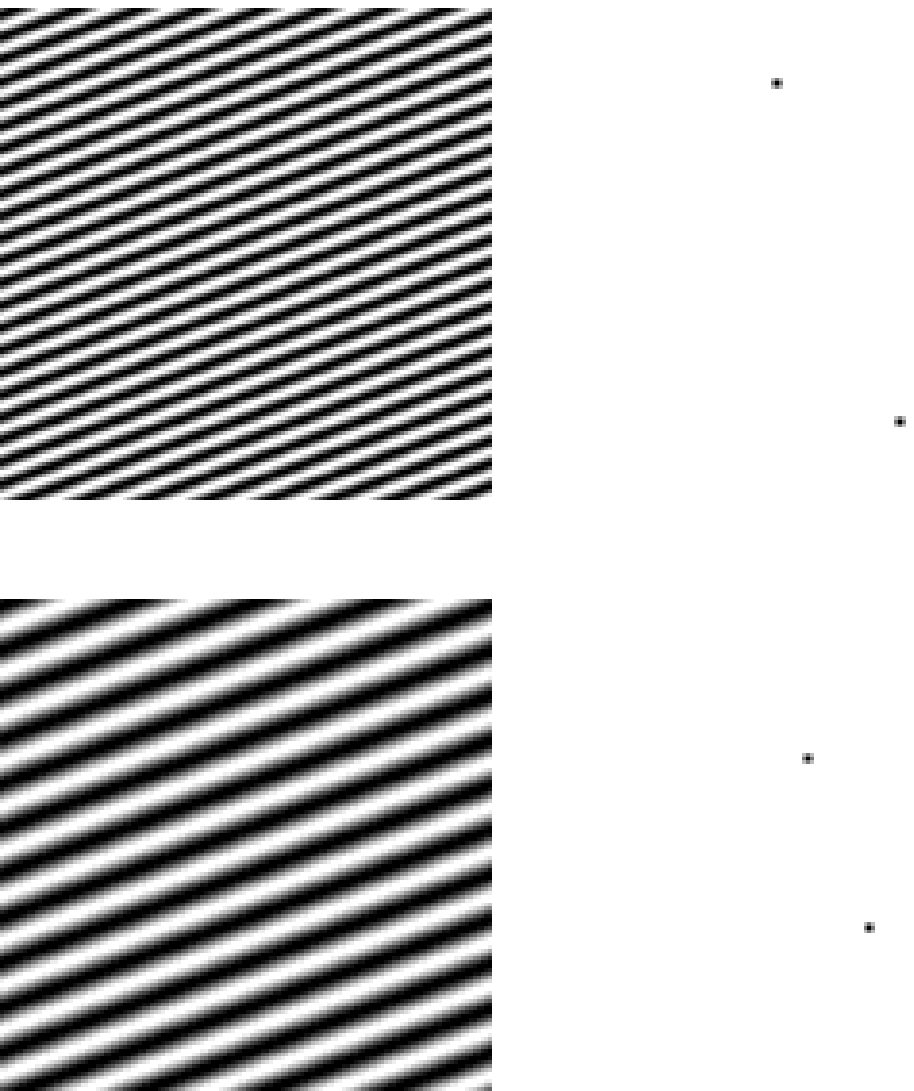}
\vspace*{1cm}
\caption{Field configuration for $\theta = -2$, $I_s=0.51$ ($I_0=10.2$) and an
integration time of $t = 10000$. From left to right and from top to bottom:
$\Re(E_x(\vec r))$ (grayscale: black=0.32, white=0.34), $|E_x(\vec k)|^2$,
$\Re(E_y(\vec r))$ (grayscale: black=-0.15, white=0.15) and $|E_y(\vec k)|^2$.}
\label{t2}
\end{figure}

For $\theta = -1.1$ we find a steady pattern in which the symmetric modes are
damped and a stripe pattern with wave number $k_a$ appears in the $\hat
y$-polarized component, while $x$-polarized component is almost homogeneous 
(in fact, a small amplitude stripe pattern of wavevector $2 k_a$ is observed in
$E_x$ which comes from the coupling between $E_x$ and $E_y$). This pattern made
out of antisymmetric modes is the same pattern that is formed when only the
antisymmetric instability is present. Overall, the electric field displays an
elliptically polarized spatial structure. Similar stationary patterns are found
for larger absolute values of the detuning. In fig. \ref{t2} we show the case
with $\theta = -2$; this value was chosen in order to induce the formation of a
square pattern, since the angle $\phi$ between symmetric and antisymmetric
vectors is $\phi = 45^\circ$ in this case, but anyway the antisymmetric stripe
pattern is formed as described.

\section{Concluding Remarks}
\label{conclusions}

We have studied the spatial polarization structures in a mean field model for a
Kerr medium close to a two-photon resonance and driven by a linearly $\hat
x$-polarized field with negative detuning. In the self-focusing case the first 
pattern that is formed as the pump intensity is increased arises from the
competition between two stationary instabilities which occur simultaneously.
This codimension two bifurcation appears naturally in the system and it is a
consequence of the form of the nonlinearities associated to the
two-photon-resonant four-wave mixing process. As this codimension two
bifurcation is not the result of the fine tuning of two parameters, as it is
usually the case, it should be much more simple to be observed. In fact there
is only one parameter, the pump, which has to be tuned. Furthermore, we still
have another free parameter, the detuning, which allows the system to form
different patters without changing the distance to the codimension two
instability threshold. 

Near the instability threshold, we have obtained the evolution equation for
the patterns arising from the interacting instabilities using a weakly
nonlinear analysis. From these analysis and from the numerical integration of
the model we have shown that we can have the following patterns: a) dynamical
structures involving an arbitrary number of symmetric and antisymmetric modes,
b) a steady state pattern formed by the superposition of a symmetric mode and
the quadratically resonant antisymmetric ones and c) an antisymmetric stripe
steady pattern in $E_y$ which is the same that would appear if only the
antisymmetric instability were present.

As a final remark we notice that an interesting peculiarity of the situation 
considered here is that the selection among these patterns can be done in a
very natural way, that is, changing the value of the detuning.

\section{Acknowledgments}

Helpful discussions with Dr. M. Santagiustina are acknowledged. This work is
supported by the European Commission through the QSTRUCT TMR network (Project
ERB FMRX-CT96-0077) and by Direcci\'on General de Ense\~nanza Superior e
Investigaci\'on Cient\'{\i}fica (Spain) Projects PB94-1167 and 
PB97-0141-C02-02. M. H. wants to acknowledge financial support from the FOMEC 
project 290, Dep. de F\'{\i}sica FCEyN, Universidad Nacional de Mar del Plata, 
Argentina.

\begin{thebibliography}{123}

\bibitem[+]{Daniel} Permanent address: Center for Nonlinear Phenomena and
Complex Systems, Universit\'e Libre de Bruxelles, Campus Plaine, Blv. du
Triomphe B.P 231, 1050 Bruxelles.

\bibitem[*]{www} Electronic address: http://www.imedea.uib.es/PhysDept/

\bibitem{Chaossolfract} {\it Nonlinear Optical Structures, Patterns, Chaos},
edited by L. A. Lugiato, special issue of Chaos, Solitons and Fractals, 
{\bf 4}, 1251 (1994) and references therein.


\bibitem{CrossHoh} M. C. Cross and P. C. Hohenberg, Rev. Mod. Phys. {\bf 65},
851 (1993).

\bibitem{lugi} L. A. Lugiato and R. Lefever, Phys. Rev.
Lett. {\bf 58}, 2209 (1987).

\bibitem{firt} W. J. Firth, A. J. Scroggie, G. S.
McDonald and L. Lugiato, Phys. Rev. A {\bf 46}, R3609 (1992).

\bibitem{geddes1} J. B. Geddes, J. V. Moloney, E.M. Wright and W.J. Firth, Opt.
 Commun. {\bf 111}, 623 (1994).

\bibitem{corrkerr} M. Hoyuelos, P. Colet and M. San Miguel, Phys. Rev. E,
{\bf 58}, 74 (1998).

\bibitem{hoyuelos} M. Hoyuelos, P. Colet, M. San Miguel and D. Walgraef, 
Phys. Rev. E, {\bf 58}, 2992 (1998).

\bibitem{Boyd92} R.W. Boyd, {\it Nonlinear Optics}, (Academic Press, San Diego,
1992).

\bibitem{Grynberg84} G. Grynberg, Opt. Commun. {\bf 48}, 432 (1984).

\bibitem{Ducloy84} M. Ducloy and D. Bloch, Phys. Rev. A {\bf 30}, 3107 (1984).

\bibitem{Malcuit88} M.S. Malcuit, D.J. Gauthier and R.W. Boyd, Opt. Lett. {\bf
13}, 663 (1988).

\bibitem{footnote1} In \cite{geddes1} $I_0$ is defined as the intensity of any
of the components of the input field, so in the equation equivalent to
(\ref{symmetricsol}) in \cite{geddes1}, $I_0$ is not divided by 2.

\bibitem{montagne} R. Montagne, E. Hern\'andez-Garc\'{\i}a, 
A. Amengual and M. San Miguel, Phys. Rev. E {\bf 56}, 151 (1997). 

\bibitem{geddes2} J. B. Geddes, R. A. Indik, J. V. Moloney and W. J. Firth, 
Phys. Rev. A {\bf 50}, 3471 (1994).

\bibitem{daniel} D.Walgraef, {\it Spatio-Temporal Pattern Formation} 
(Springer-Verlag, New York, 1996).

\bibitem{haken} M. Bestehorn and H. Haken,
Phys. Rev A {\bf 42}, 7195 (1990).


\end{thebibliography}
\end{document}